
\input phyzzx.tex

  \def\tr{{\hbox{\rm Tr}}}

\tolerance=500000 \overfullrule=0pt
 \def\np{Nucl. Phys.}
\def\pl{Phys. Lett.} \def\pre{Phys. Rep.} 
  \def\cmp{Comm. Math. Phys.}
\def\ijmp{Int. J. Mod. Phys.}      \def\topo{Topology}   \def\jmp{J. Math. Phys.} \def\jgp{J.
Geom. Phys.} \def\jdg{J. Diff. Geom.}
\def\plms{Proc. London Math. Soc.}
\def\mrl{Math. Research Lett.}

    \def\ker{{\hbox{\rm Ker}}} \def\tr{{\hbox{\rm Tr}}}

\def\too{\longrightarrow}

\tolerance=500000 \overfullrule=0pt

\pubnum={US-FT-4/95\cr
hepth@xxx/9504010}
\date={April, 1995}
\pubtype={}
\titlepage \title{NON-ABELIAN MONOPOLES ON FOUR-MANIFOLDS}
\author{J.M.F. Labastida\foot{E-mail: LABASTIDA@GAES.USC.ES} and  M.
Mari\~no} \address{Departamento de F\'\i sica de Part\'\i culas\break
Universidade de Santiago\break E-15706 Santiago de Compostela, Spain}
\abstract{We present a non-abelian generalization of Witten
monopole equations and we analyze the associated moduli problem, which can
be regarded as a generalization of Donaldson theory. The moduli space of
solutions for $SU(2)$ monopoles on K\"ahler manifolds is discussed. We also
construct, using the Mathai-Quillen formalism, the topological quantum field
theory corresponding to the new moduli problem. This theory involves the
coupling of topological Yang-Mills theory to topological matter in four
dimensions.}

\endpage
\pagenumber=1

\chapter{Introduction.}

Topological quantum field theory has shown to be a very fruitful arena in
both, physics and mathematics. Since the formulation of Donaldson theory
\REF\donfirst{S.K. Donaldson\journal\jdg&18(83)279}
\REF\don{S.K. Donaldson\journal\topo&29(90)257}
\REF\donkron{S.K. Donaldson and P.B. Kronheimer,
{\it The Geometry of Four-Manifolds},
Oxford Mathematical Monographs, 1990} [\donfirst,\don,\donkron] in
the language of quantum field theory
\REF\tqft{E. Witten\journal\cmp&117(88)353} [\tqft] many other interesting
aspects of topology and algebraic geometry have been reached. In recent
years, the physical and mathematical aspects of topological quantum field
theory have benefited from each other. On the one hand, using insight from
topological aspects related to Donaldson-Witten theory \REF\wjmp{E.
Witten\journal\jmp&35(94)5101} [\wjmp], theories with $N=4$
\REF\wv{C. Vafa and E.
Witten \journal\np&B431(94)3 } [\wv]
\REF\ws{N. Seiberg and E. Witten
\journal\np&B426(94)19 \journal\np&B431(94)484 }
and $N=2$ [\ws] supersymmetry in four dimensions have been
solved in the infrared; on the other hand, these developments in physical
theories have been used to show that Donaldson-Witten theory itself can be
mapped to a simpler theory involving abelain monopoles\REF\mfm{E.
Witten\journal\mrl&1(94)769} [\mfm].

Donaldson theory involves the study of certain differential forms on
the moduli space of self-dual gauge connections. This theory, initially
formulated for $SU(2)$, can be reformulated in terms of quantum
field theory [\tqft] and generalized to other gauge groups.
The theory is intrinsically non-abelian.
In [\mfm] Witten showed that Donaldson theory with gauge group
$SU(2)$ is equivalent to a new moduli problem which involves an abelian
connection coupled to matter in a pair of monopole equations.
The topological quantum field theory associated to this new moduli space
has been recently constructed in
\REF\toplag{J.M.F. Labastida and M. Mari\~no, A topological Lagrangian
for monopoles on
four-manifolds, hep-th/9503105} [\toplag] using the Mathai-Quillen formalism.
The resulting theory turned out to be an abelian Donaldson-Witten theory
coupled to a twisted version of the $N=2$ supersymmetric hypermultiplet
\REF\fay{P.
Fayet\journal\np&B113(76)135}
\REF\son{M.F. Sohnius\journal\np&B138(78)109}
\REF\hst{P.S. Howe, K.S. Stelle
and P.K. Townsend\journal\np&B214(83)519} [\fay,\son,\hst].
A non-abelian version of this model, associated though to a simpler
moduli problem, was presented in
\REF\alab{M. Alvarez and J.M.F. Labastida\journal\pl&B315(93)251}
\REF\alabas{M. Alvarez and J.M.F. Labastida\journal\np&B437(95)356}
[\alab,\alabas]. Related topological quantum field theories have been
analyzed in  \REF\ans{D. Anselmi
and P. Fre\journal\np&B392(93)401\journal\np&B404(93)288
\journal\np&B416(94)255} [\ans], and their connection to [\mfm] has
 been
indicated in
\REF\ansdos{D. Anselmi
and P. Fre\journal\pl&B347(95)247} [\ansdos].

In this paper we present the non-abelian version of the monopole equations
proposed in [\mfm], we study the moduli problem and the associated moduli
space for the simple case of $SU(2)$ and matter fields in the fundamental
representation, and we construct the corresponding topological quantum field
theory action using the Mathai-Quillen formalism. The generalization of the
monopole equations involve the following data: a four-dimensional spin
manifold $X$ endowed with a metric $g$, a gauge group $G$ and a
representation $R$. The data are therefore enlarged with respect to the
ordinary
Donaldson-Witten theory by the presence of matter in a representation $R$. This
is an indication that the set of topological quantities associated to this new
theory might be richer than in the ordinary case. It is interesting  to notice
that the step given in [\mfm] to map Donaldson theory to a simpler one is now
utilized to enlarge the original theory. In fact we will show that the
conditions to have a well-defined moduli problem are the same than in the
Donaldson case, and the non-abelian monopole theory appears thus as a
natural generalization of the Donaldson theory. We will argue that the
new moduli space contains the moduli space of anti self-dual connections and
in addition new branches of solutions which in the $SU(2)$ case are
similar to the ones that appear in the abelian theory.

The analysis carried out in this work although often is particularized for
simple cases can in principle be extended to the general situation.
We have concentrated in the specific cases working along the lines of
[\don] and [\mfm] but presumably similar arguments can be used in general.

The paper is organized as follows. In sect. 2 we present the monopole
equations and we analyze the corresponding moduli problem. In sect. 3 the
previous analysis is extended for the case of $SU(2)$ monopoles on K\" ahler
manifolds, with the matter fields in the fundamental representation. In sect.
4 the non-abelian topological action is constructed using the Mathai-Quillen
formalism. Finally, in sect. 5 we state our conclusions. An appendix contains
the spinor conventions used in the paper.

\chapter{Non-abelian monopole equations.} Let $X$ be an oriented, closed
four-manifold endowed with a Riemannian structure given by a metric $g$. We
will restrict ourselves to spin manifolds, although the generalization to
arbitrary manifolds can be done using a ${\rm Spin}_c$ structure. We will
denote the positive and negative chirality spin bundles on $X$ by $S^{+}$ and
$S^{-}$, respectively. In [\mfm], Witten introduced a new moduli problem
involving an abelian Yang-Mills field, associated to a $U(1)$ complex line
bundle $L$, coupled to a spinor field of positive chirality in a pair of
``monopole equations":
$$
\eqalign{ &F^{+}_{\alpha \beta}+{i\over 2}
{\overline M}_{(\alpha}  M_{\beta )}=0, \cr & D_{\alpha {\dot \alpha}
}M^{\alpha}=0,\cr}
\eqn\ab
$$
where $D_{\alpha {\dot \alpha}}$ is the Dirac
operator and $F^{+}_{\alpha \beta}$ the self-dual part of the gauge
field-strength (see eq. (A31,A40)
in the appendix for the conventions used). This moduli
problem turns out to be equivalent to the $SU(2)$ Donaldson theory on $X$. In
principle a non-abelian generalization of these monopole equations would give
rise to a generalization of Donaldson theory which from the physical point of
view corresponds to a coupling of a topological Yang-Mills theory to
topological matter in four dimensions [\alab, \alabas]. As we will see, the
non-abelian monopole equations share many properties of Donaldson theory as
well as of the abelian theory proposed in [\mfm].
\REF\aj{M.F. Atiyah and L. Jeffrey\journal\jgp&7(90)119}
\REF\cmrym{S. Cordes, G. Moore and S. Rangoolam,
``Large N 2D Yang-Mills Theory and
Topological String Theory", hep-th/9402107, 1994}
\REF\cmrlect{S. Cordes, G. Moore and S.
Rangoolam, Proceedings of the 1994 Les Houches Summer School,
 hep-th/9411210}

Before going on with the non-abelian generalization of \ab, it is important
to recall the topological framework for the field-theory approach to moduli
problems in the context of the Mathai-Quillen formalism [\aj,\cmrlect]: given a
(infinite-dimensional) field space ${\cal M}$ and a vector bundle over ${\cal
M}$, ${\cal V}$, the basic equations of the problem are defined as sections of
this vector bundle. Let us denote generically these sections as $s:{\cal M}
\too
{\cal V}$. In the situations in which there is a gauge symmetry, as will be the
case under consideration, one  has to take into account the action of a group
${\cal G}$ on both, the manifold ${\cal M}$ and the vector bundle. This is done
``dividing by ${\cal G}$" which implies that the section $s$ must be taken to
be
gauge-equivariant and hence one must consider the associated section ${\hat
s}:{\cal M}/{\cal G} \too {\cal V} /{\cal G}$.

The basic topological invariant
associated to the moduli problem is the Euler characteristic of the bundle
${\cal V}$, which can be obtained,
as in the finite-dimensional case, integrating the
pullback under $s$ of its Thom class on ${\cal M}$. In the situations with
gauge symmetries the interest resides in the computation of this class for
the corresponding quotient bundle. \REF\mq{V. Mathai and D.
Quillen\journal\topo&25(86) 85}The computation of the Thom class involves the
construction of the Mathai-Quillen form [\mq] which leads to the topological
quantum field theory associated to the moduli problem. The Mathai-Quillen
form for the non-abelian monopole theory will be obtained in sect. 4. We will
be interested in the special situation in which the vector bundle ${\cal V}$
is trivial and can be written in the form ${\cal V} = {\cal M} \times {\cal
F}$, where ${\cal F}$ is the fibre on which a ${\cal G}$-invariant metric is
defined. Considering  the moduli space ${\cal M}$ as a principal bundle with
group ${\cal G}$ the quotient bundle is the associated vector bundle ${\cal
E}={\cal M} {\times _{\cal G}} {\cal F}$. This is the situation common to
Donaldson-Witten theory and monopoles on four-manifolds as described in [\aj]
and [\toplag] respectively. A discussion for the case in which  there are
non-trivial vector bundles can be found in [\cmrym, \cmrlect].

In Witten's monopole theory, as it is discussed
in [\toplag], the geometrical data are a four-manifold $X$ and a complex line
bundle $L$ over $X$, and the field space is ${\cal M}={\cal A} \times \Gamma
(X, S^{+} \otimes L)$, where ${\cal A}$ is the moduli space of $U(1)$ abelian
connections on $L$, and $\Gamma (X, S^{+} \otimes L)$ are the sections of the
product bundle $S^{+} \otimes L$, {\it i.e.}, positive chirality spinors
taking values in $L$.  The vector bundle over ${\cal M}$ is a trivial one
with fibre ${\cal F}=\Omega^{2,+} (X) \oplus \Gamma (X, S^{-} \otimes L)$,
where the first factor denotes the self-dual differential forms of degree
$2$ on $X$. The equations \ab\ define a moduli space which is the zero locus
of a section of this bundle. The group ${\cal G}$ is the group of gauge
transformations of the principal $U(1)$-bundle associated to the connection
$A$.

The obvious way to construct the non-abelian moduli problem consists of
considering, instead of a complex line bundle, a principal fibre bundle $P$
with some compact, connected, simple Lie group $G$. The Lie algebra of $G$
will be denoted by ${\bf g}$. For the matter part we need an associated
vector bundle $E$ to the principal bundle $P$ by  means of a representation $R$
of the Lie group $G$. The field space is ${\cal M}={\cal A} \times \Gamma
(X, S^{+} \otimes E)$, where ${\cal A}$ is now the moduli space
of $G$-connections on $E$, and the spinors take values now in this
representation space.  The vector bundle over ${\cal M}$ is again a trivial
one with fibre ${\cal F}=\Omega^{2,+} (X, {\bf g}_E) \oplus \Gamma (X, S^{-}
\otimes E)$, and the self-dual differential forms take values in the
representation of the Lie algebra of $G$ associated to $R$, ${\bf g}_E$. The
group
${\cal G}$ is the group of gauge transformations of the bundle $E$, and its
action on the moduli space is given locally by:
$$
\eqalign{ &g^{*}(A_\mu)=-igd_\mu g^{-1}+gA_\mu g^{-1},\cr
&g^{*}(M_{\alpha})=g M_{\alpha},\cr }   \eqn\gauge
$$
where $M \in \Gamma (X, S^{+} \otimes E)$ and $g$ takes values in the group
$G$ in the representation $R$. Notice that in terms of the covariant
derivative $d_A=d+i[A,\;\;]$ the infinitesimal form of the transformations
\gauge\ becomes, after considering
$g=\exp(i\phi)$:
$$
\eqalign{
\delta A = -d_A \phi,\cr
\delta M_\alpha = i\phi M.\cr}
\eqn\gaugedos
$$

The group of gauge transformations also acts on the fibre ${\cal F}$, but we
must use $g^{-1}$, as  the construction of an associated vector bundle
imposes. The Lie algebra of the group
${\cal G}$ is Lie$({\cal G})=\Omega^0(X,{\bf g}_E)$. The tangent space to the
moduli space at the point $(A,M)$ is just
$T_{(A,M)}{\cal M}=T_{A}{\cal A} \oplus T_{M}\Gamma (X, S^{+} \otimes
E)=\Omega^{1}(X,{\bf g}_E)
\oplus \Gamma (X, S^{+} \otimes E)$, for $\Gamma (X, S^{+}
\otimes E)$ is a vector space.
We can define a gauge-invariant Riemannian  metric on ${\cal M}$ given by:
$$
\langle (\psi, \mu), (\theta, \nu) \rangle=\int_{X} \tr(\psi
\wedge *\theta) +{1 \over 2} \int_{X} e ({\bar \mu}^{\alpha i}
\nu_{\alpha}^i+ \mu_{\alpha}^i {\bar \nu}^{\alpha i}),
\eqn\pina
$$
where $e=\sqrt g$. The spinor notation used in this paper is conveniently
compiled in the appendix. An analogous expression gives the inner product on
the fibre ${\cal F}$. The Lie algebra of the gauge group of transformations
${\rm Lie}({\cal G})$ is also endowed with a metric given, as in \pina, by
the trace and the inner product on the space of zero-forms.

Within this general framework it is easy to write out explicitly the
non-abelian monopole equations for the cases in which Donaldson theory has
been proved more useful: the $SU(N)$ and $SO(N)$ cases. The
non-abelian monopole equations in this case are simply:
$$
\eqalign{ &F_{\alpha\beta}^{+ij}+{i \over 2}({\overline
M}^j_{(\alpha} M^i_{\beta )}-{\delta^{ij} \over d_R}{\overline M}^k_{(\alpha}
M^k_{\beta )})=0,\,\,\,\,\ {\rm for} \,\,\, SU(N), \cr
&F_{\alpha\beta}^{+ij}+{i \over 2}{\overline M}^{[j}_{(\alpha}
M^{i]}_{\beta )}=0,\,\,\,\,\ {\rm for}\,\,\, SO(N),\cr}
\eqn\nonab
$$
where $F_{\alpha\beta}^{+ij}$ are in a  representation $R$,
\ie, $F_{\alpha\beta}^{+ij}=F_{\alpha \beta}^{+a} (T^a)^{ij}$
being $T^a$ the generators of the Lie algebra taken in the
representation $R$.
In the first equation of \nonab\ (and similar ones in this paper), a
sum in the repeated index $k$ is understood, and
$d_R$ denotes the dimension of the representation $R$. The other
monopole equation is simply the Dirac equation with the Dirac operator
coupled to the gauge connection in the corresponding representation.

The structure of equations \nonab\ for other groups possesses a similar
structure. It can be written in a compact form as:
$$
F_{\alpha\beta}^{+a}+{i \over 2}{\overline
M}_{(\alpha} (T^{a}) M_{\beta )}=0,
\eqn\compacto
$$
where ${\overline M}_{(\alpha} (T^{a}) M_{\beta )}$ is shortened
form for ${\overline M}^i_{(\alpha} (T^{a})^{ij} M^j_{\beta )}$
(this convention will be used throughout the paper).
The expressions in \nonab\ are obtained from this equation after contracting
it with $T^a$ and using the fact that the normalization of the generators
can be chosen such that for the representation $R$ one has $(T^a)^{ij}
(T^a)^{kl} = \delta^{il}\delta^{jk}-{1\over d_R}\delta^{ij} \delta^{kl}$ for
$SU(N)$ and $(T^a)^{ij} (T^a)^{kl} =\delta^{i[l}\delta^{k]j}$ for $SO(N)$.
In the rest of this paper we will mainly focus on the $SU(N)$ case. The
generalization for $SO(N)$ and for other groups is straightforward.

When writing the section of the bundle
${\cal V}$ from \nonab\ it will be useful to rescale the first monopole
equation by the factor $1/\sqrt{2}$, as in the abelian theory. The section
reads therefore, in the $SU(N)$ case:
$$
s(A,M)=\Big({1 \over \sqrt 2} \big(F^{+ ij}_{\alpha \beta}+{i\over 2}
({\overline M}_{(\alpha}^j  M_{\beta )}^i-{\delta^{ij}\over d_R} {\overline
M}_{(\alpha}^k  M_{\beta )}^k)\big), (D_{\alpha {\dot \alpha}
}M^{\alpha})^i\Big).
\eqn\section
$$

A first step to understand the structure of the moduli space of solutions
to the non-abelian monopole equations modulo gauge transformations is to
construct the associated instanton deformation complex, which allows one to
compute (under certain assumptions) the dimension of the tangent space to
this moduli space (the virtual dimension of the moduli space). For this we
need an explicit construction of the gauge orbits, which are given by the
vertical tangent space on the principal bundle with group ${\cal G}$. This
space is the image of a map from the Lie algebra of
the group
${\cal G}$ to the tangent space to ${\cal
M}$,
$$ C:{\rm Lie}({\cal G})  \too T{\cal M},
\eqn\ope
$$
which can be obtained from \gaugedos\ and reads:
$$
C(\phi)=(-d_{A}\phi ,i\phi^{ij} M^j) \in \Omega^{1}(X,{\bf g}_E) \oplus \Gamma
(X, S^{+}
\otimes E),\,\,\,\,\ \phi \in \Omega^{0}(X,{\bf g}_E).
\eqn\flor
$$
Using the metrics in \pina\ and the analogous one on the fibre, we can compute
the adjoint operator $C^{\dagger}$ which will be needed later to obtain the
topological lagrangian of the theory. Let us consider $(\psi,\mu) \in
T_{(A,M)}{\cal M}= \Omega^{1}(X,{\bf g}_E) \oplus \Gamma (X, S^{+} \otimes
E)$. One finds,
$$
C^{\dagger}(\psi,\mu)^{ij}=-(d_A^{*}\psi)^{ij} + {i \over 2}\big({\bar
\mu}^{\alpha j}M_{\alpha}^i-{\overline M}^{\alpha
j}\mu_{\alpha}^i-{\delta^{ij}\over d_R} ({\bar \mu}^{\alpha
k}M_{\alpha}^k-{\overline M}^{\alpha k}\mu_{\alpha}^k)\big) \in
\Omega^{0}(X,{\bf g}_E). \eqn\marga
$$

We also need the linearization of the non-abelian monopole equations, which
can be understood as a map $ds:T_{(A,M)}{\cal M} \too {\cal F}$. The result
is:
$$
\eqalign{ ds(\psi,\mu) = & \bigg( {1 \over \sqrt
2}\Big(\big(p^+(d_A\psi)\big)_{\alpha \beta}^{ij}+{i\over 2}\big({\overline
M}_{(\alpha}^j \mu_{\beta )}^i+{\bar \mu}_{(\alpha}^j M_{\beta )}^i
-{\delta^{ij}\over d_R}({\overline M}_{(\alpha}^k \mu_{\beta )}^k+{\bar
\mu}_{(\alpha}^k M_{\beta )}^k)\big)\Big),\cr &\,\,\,\,\,\,\,\,\,\,\,
(D_{\alpha {\dot \alpha} } \mu^{\alpha})^i+i\psi_{\alpha {\dot
\alpha}}^{ij}M^{\alpha j}  \bigg), \cr} \eqn\disco
$$
where $p^+$ is the projector defined in the appendix (eq. (A40)). The maps $ds$
and $C$ fit into the instanton deformation complex:
$$
0 \too \Omega^{0}(X,{\bf g}_E)
\buildrel C \over \too \Omega^{1}(X,{\bf g}_E) \oplus \Gamma (X, S^{+} \otimes
E) \buildrel ds \over \too
 \Omega^{2,+} (X,{\bf g}_E) \oplus \Gamma (X, S^{-} \otimes E) \too 0.
\eqn\file
$$
The index of this complex can be computed dropping terms of order zero in the
operators $C$ and $ds$ (as their leading symbol is not changed). In this way,
the complex \file\ splits into the complex associated to the anti-self-dual
(ASD) connections of Donaldson theory:
$$
0 \too \Omega^{0}(X,{\bf g}_E)
\buildrel d_A \over \too \Omega^{1}(X,{\bf g}_E)
\buildrel p^{+}d_A \over \too
 \Omega^{2,+} (X,{\bf g}_E)  \too 0.
\eqn\asd
$$
 and the complex of the twisted Dirac operator. The index will be simply the
sum of the virtual dimension of the moduli space of ASD instantons, ${\cal
M}_{\rm ASD}$, and of twice the index of the twisted Dirac complex (for we
are considering $S^{+} \otimes E$,
$S^{-} \otimes E$ as real vector bundles, in order to obtain the real
dimension of the moduli space). This is easily computed and gives:
$$
{\rm index}\,\ D=\int_X {\rm ch}(E) {\hat A}(X)=-{d_{R} \over 8}\sigma
-c_2(E),
\eqn\indice
$$
where $\sigma$ is the signature of the four-manifold $X$, which is given,
according to the Hirzebruch signature formula, by $\sigma=p_1(X)/3$, and
$c_2(E)$ is the second Chern class of the representation bundle. The virtual
dimension of the moduli space for the non-abelian theory,
${\cal M}_{\rm NA}$, in the
$SU(N)$ case, is
thus
$$
{\rm dim} \,\ {\cal M}_{\rm NA}={\rm dim} \,\ {\cal M}_{\rm ASD}+2{\rm
index}\,\ D=(4N-2)c_2(E)-{N^2-1 \over 2}(\chi +\sigma)-{d_{R} \over 4}\sigma,
\eqn\dim$$
where $\chi$ is the Euler characteristic of $X$. The generalization of this
expression to other gauge groups is straightforward.

There are two points that are important in order to understand the
non-abelian monopole equations and the possibility of extracting some new
topological information from them. The basic topological invariant associated
to the non-abelian equations \nonab\ is the Euler characteristic of the bundle
${\cal E}$ and can be interpreted as the partition function of the
corresponding topological theory (which we will construct in sect. 4). This
partition function is defined for ${\rm dim} \,\ {\cal M}_{\rm NA}=0$, {\it
i.e.}, when the moduli space of solutions of the monopole equations consists
of a finite set of points. On the tangent space to each of these points we can
consider the elliptic operator associated to the complex \file,
$T=C^{\dagger}\oplus ds$, and the sign of the determinant of this operator.
Using standard arguments [\tqft] one can show that the partition function of
the
theory is given by the sum over the set of solutions of the monopole
equations of the signs of ${\rm det}\,\ T$ (this is a infinite dimensional
version of the Poincar\'e-Hopf theorem [\cmrlect, \wv]). But to have a
well-defined topological quantity we need the determinant bundle of $T$ to be
a trivial one. From a field-theory point of view, this is equivalent to
require that the corresponding topological field theory does not have global
anomalies. Now, recall that the operator $T$ can be deformed to a direct sum
of the Dirac operator and the elliptic operator for the complex of ASD
connections \asd. The determinant line bundle of the Dirac operator, when
this is regarded as a real operator, has a natural trivialization coming from
its underlying complex structure. Therefore one must prove the triviality of
the determinant bundle of the operator $\delta_A=d_A^{*}\oplus p^{+}d_A$
coming from \asd. But this is in fact guaranteed by Donaldson theory
\REF\donor{S.K. Donaldson\journal\jdg&26(87)397}
[\donor,\donkron] and it is equivalent to the orientability of the moduli
space of irreducible connections with $H^{2}_A={\rm coker}\,\ p^{+}d_A=0$
(for such connections, ${\rm coker}\,\  \delta_A=0$ and therefore the
determinant line bundle of $\delta_A$ coincides with $\Lambda^{\rm max}\ker
\delta_A=\Lambda^{\rm max}T_A {\cal M}_{\rm ASD}$).

The second important question which arises in order to have a well-defined
moduli problem is whether or not the group of gauge transformations has a
free action on the space of solutions to the non-abelian monopole equations.
This is rather obvious when we look at the partition function as the Euler
class of the bundle ${\cal E}$: if the action of the gauge group has fixed
points, we will have singularities when making the quotient  by ${\cal G}$
and we will not be able to get rid of the gauge degrees of freedom. From the
field-theory point of view \REF\witcoh{E. Witten\journal\ijmp&A6(91)2775 }
[\witcoh] a non-free action of the gauge group gives a moduli space larger
than the space to which we want to localize the path integral of the
corresponding topological field theory. Notice that, for the non-abelian
case, the only way to have a fixed point in the space of solutions to the
monopole equations is to have, as in the abelian case, a solution with $M=0$.
In this case, the monopole equations reduce to the equation which defines an
ASD connection. In fact, the non-abelian monopole equations have always the
solution $M=0$, $F^{+}=0$, and therefore ${\cal M}_{\rm ASD} \subset {\cal
M}_{\rm NA}$. A non-free action of the gauge group can only be possible in
the subset of ASD connections, {\it i.e.}, when we have reducible ASD
connections. The conditions for a free action are thus the same than in
Donaldson theory. This analysis and the one we did for establishing the
triviality of the determinant line bundle show that the non-abelian monopole
theory appears as a rather natural generalization of Donaldson theory: the
moduli space of solutions contains ${\cal M}_{\rm ASD}$ as a subset, and the
conditions for having a well defined moduli problem are essentially the same.
In this way it seems that the coupling of non-abelian topological Yang-Mills
theory to topological matter in four dimensions given by the non-abelian
monopole equations could provide an adequate extension of the Donaldson
framework. Indeed, we will try to argue in the next section that the moduli
space of solutions to the non-abelian equations has in principle a richer
structure than ${\cal M}_{\rm ASD}$.

Another aspect of the relation of the non-abelian monopole theory to
Donaldson theory is the following. In the abelian case, Witten showed [\mfm],
making use of vanishing theorems, that there are only a finite number of
isomorphism classes of line bundles for which the moduli space of solutions
has a positive or zero virtual dimension. In the $SU(N)$ case, as for the
abelian monopoles, vanishing theorems are obtained by computing the squared
norm of the section \section\ using the natural Riemannian metric on the
fibre. Let us carry out the corresponding analysis for the non-abelian case.

Taking into account the Weitzenb\"ock formula (see eq. (A44) in the
appendix):
$$
D_{\alpha\dot\alpha} D_\beta{}^{\dot\alpha} M^{\beta i}= (g^{\mu \nu}D_{\mu}
D_{\nu}-{1\over 4} R) M^{i}_\alpha + i F^{+ ij}_{\alpha\beta} M^{j\beta},
\eqn\weitz
$$
being $R$ the scalar curvature on $X$, one finds:
$$
\eqalign{ |s(A,M)|^{2}&= \int _{X} e {D}_{\alpha\dot\alpha} {\overline
M}^{\alpha} D_{\beta}{}^{\dot \alpha }M^{\beta} \cr  &{\hbox{\hskip-2cm}}+{1
\over 2}\int _{X}e \big(F^{+\alpha \beta ji} +{i\over 2}({\overline
M}^{i(\alpha}  M^{\beta )j}-{\delta^{ij}\over d_R}{\overline M}^{k(\alpha}
M^{\beta )k})\big)  \big(F^{+ ij}_{\alpha \beta} +{i\over 2}({\overline
M}^{j}_{(\alpha} M^{i}_{\beta )}-{\delta^{ij}\over d_R}{\overline
M}^{k}_{(\alpha} M^{k}_{\beta )})\big) \cr & =\int_{X} e \big[
g^{\mu\nu}D_\mu\overline M^\alpha D_\nu M_\alpha + {1\over 4} R \overline
M^\alpha M_\alpha +{1\over 2} \tr (F^{+\alpha\beta} F_{\alpha\beta}^+) \cr
&\,\,\,\,\,\,\,\,\,\,\,\,\,\,\,\,\,\,\,\,\, -{1\over 8} (\overline
M^{i(\alpha} M^{\beta)j} \overline M_{(\alpha}^j M_{\beta)}^i -{1\over d_R}
\overline M^{i(\alpha} M^{\beta)i} \overline M_{(\alpha}^j
M_{\beta)}^j\big)]. \cr}
\eqn\melon
$$
After using the fact that for
$SU(N)$ the normalization of the generators $T^a$ can be chosen in such a
way that for the representation $R$ $(T^a)^{ij} (T^a)^{kl} =
\delta^{il}\delta^{jk}-{1\over d_R}\delta^{ij} \delta^{kl}$, the last term in
\melon\ can be written as:
$$
-{1\over 8}(\overline
M^{(\alpha} T^a M^{\beta)}) (\overline M_{(\alpha} T^a M_{\beta)}),
\eqn\current
$$
where a sum over $a$ is must be understood.
Notice that if one denotes the components of
$M_\alpha^i$ by $M_\alpha^i=(a^i,b^i)$, this term is in fact,
$$
{1\over 4}\Big( (1-{1 \over d_R})\big(\sum_{i}(|a^i|^2+|b^i|^2)\big)^2 +(1+{2
\over d_R})\sum_{ij}|a^{[i}b^{j]}|^2 \Big),
\eqn\melondos
$$
and therefore it is positive definite. The factor $i\overline M^\alpha
F^+_{\alpha\beta} M^\beta$ has cancelled in the sum, and then each term in the
second expression for $|s(A,M)|^2$ in \melon\ is positive definite except the
one involving the scalar curvature. This was the reason of choosing  the
factor ${1/ {\sqrt 2}}$ in \section.

The advantage of the form \melon, which will become the
bosonic sector of the topological action is that, as discussed in sect. 3, one
can apply vanishing theorems which improve  the analysis of the space of
solutions of the monopole equations as in [\mfm,\wv]. As in the abelian case,
we can get from this expression an upper bound for the squared norm of the
self-dual part of the curvature on solutions of the monopole equations:
$$
I^{+}=\int _{X}e F_a^{+\alpha\beta} F_{a,\alpha\beta}^+ \le {1 \over
8(1-1/d_R)}\int_{X}eR^{2}
\eqn\melontres
$$
But, in contrast to the abelian case, we cannot find an upper bound for the
anti self-dual part $I^{-}$ when we impose that dimension of the moduli
space \dim\ to be greater than or equal to zero. This is so because a theory
of $SU(N)$ connections involves the instanton number $c_2(E)$, which equals
$I^{-}-I^{+}$. Therefore the non-abelian theory is in this respect like
Donaldson theory.

\chapter{$SU(2)$ monopoles on K\"ahler manifolds.} In
this section we will analyze in more detail the non-abelian monopole
equations on a compact K\"ahler manifold $X$ and for the case in which the
gauge group is $SU(2)$, following the procedure of [\mfm]. We will show that
many of the characteristics of the abelian case are shared by the non-abelian
equations, and this will allow us to propose a more concrete picture of
${\cal M}_{\rm NA}$ in this case. In particular we want to argue that this
moduli space is in fact ``larger" than
${\cal M}_{\rm ASD}$, and therefore that the non-abelian theory can give a
different kind of topological information which may be useful for studying
the geometry of four-manifolds.

On a K\"ahler manifold the spinor bundle $S^{+}$ splits into $K^{1/2}\oplus
K^{-1/2}$, where $K^{1/2}$ is a square root of the canonical bundle $K$. Let
$E$ be the vector bundle associated to the fundamental representation of
$SU(2)$, and denote by $\alpha=(\alpha^1, \alpha^2)$ and $-i{\overline
\beta}=(-i{\overline \beta}^1,-i{\overline \beta}^2)$ the components of
$M^i_\alpha$ in $K^{1/2}\otimes E$ and $K^{-1/2}\otimes E$, respectively. If
we denote by $\omega$ the K\"ahler form on $X$, we have the decomposition of
self-dual forms $\Omega^{+}=\Omega^{2,0}\oplus \Omega^{0,2}\oplus
\Omega^{0}\omega$. According to this decomposition we can write the first
$SU(2)$ monopole equation as:
$$
\eqalign{
F_{2,0}^{ij}=&\alpha^i \beta^j-{1 \over 2}\delta^{ij}\alpha^k \beta^k,\cr
F_{\omega}^{ij}=&-{\omega \over 2}\Big( \alpha^i {\overline
\alpha}^j-{\overline \beta}^i \beta^j-{\delta^{ij} \over
2}(|\alpha^k|^2-|\beta^k|^2) \Big),\cr
F_{0,2}^{ij}=&{\overline \alpha}^j {\overline \beta}^i-{1 \over
2}\delta^{ij}{\overline \alpha}^k
{\overline \beta}^k.\cr}
\eqn\kal$$
Now we can use expression \melon\ to obtain vanishing results for the
solutions of \kal, as in [\mfm,\wv]. Suppose $(A,\alpha,\beta)$ is a solution
to \kal, and hence \melon\ vanishes. Then $(A,\alpha, -\beta)$ makes \melon\
vanish too, and we obtain another solution to \kal. Therefore, any solution of
these equations verifies:
$$
F_{2,0}^{ij}=F_{0,2}^{ij}=0.
\eqn\hol
$$
This tells us that the connection $A$ endows $E$ with the structure of a
holomorphic bundle, as it happens in Donaldson theory for ASD connections and
in the abelian theory.

The most general solution to the equation
$$
\alpha^i \beta^j-{1 \over 2}\delta^{ij}\alpha^k \beta^k=0,
\eqn\lisa
$$
is $\alpha \not= 0$, $\beta = 0$ or $\alpha=0$, $\beta \not= 0$ (with
$\alpha$, $\beta$ understood as vectors). Of course we also have the solution
$\alpha =\beta = 0$, which corresponds to an ASD instanton. We want to
consider the first kind of solutions. Suppose $\alpha \not= 0$, $\beta = 0$.
The Dirac equation for this kind of solution is simply ${\overline
\partial}_{A}\alpha=0$, with ${\overline \partial}_{A}$ the twisted Dolbeault
operator on $E$. As $A$ defines a holomorphic structure on $E$, according to
\hol, the Dirac equation simply tells us that $\alpha$ is a holomorphic
section of the bundle $K^{1/2}\otimes E$. For $\alpha=0$, $\beta \not= 0$ we
have the symmetric situation, with $\beta$ a holomorphic section of the bundle
$K^{1/2}\otimes {\overline E}$. Now we want to consider the second equation of
\kal. For this we will use some techniques of symplectic geometry which have
been proved to be useful both in Donaldson theory [\donkron] and in the
abelian case [\mfm]. Suppose again we are in the case $\alpha \not= 0$,
$\beta = 0$. We define a symplectic structure on ${\cal M}_{\beta =0}={\cal
A}\times \Gamma (X,K^{1/2}\otimes E)$ according to: $$
\Omega ((\psi, \mu), (\theta, \nu) )=\int_{X}  {\rm Tr} (\psi \wedge
\theta)\wedge \omega -{i
\over 2} \int_{X} \omega \wedge \omega ({\bar \mu}^i \nu^{i}- \mu
^{i} {\bar \nu}^i),
\eqn\sym
$$
where $\psi$, $\theta$ are in $\Omega^{1}(X,{\bf g}_E)$ and $\mu$, $\nu$ in
$\Gamma (X,K^{1/2}\otimes E)$. This symplectic form is obviously
preserved by the action of the group of gauge transformations. We consider
$\Omega^{4}(X,{\bf g}_E)$ as the dual of
${\rm Lie}({\cal G})=\Omega^{0}(X,{\bf g}_E)$, and the pairing is given by the
integration over $X$ of the trace of the wedge product. A moment map for the
action  of the group of gauge transformations is a map,
$$
m:{\cal M}_{\beta =0} \too \Omega^{4}(X,{\bf g}_E),
\eqn\mapa
$$
verifying:
$$ \langle (dm)_{(A, \alpha)}(\psi, \mu), \phi \rangle = \Omega ((\psi, \mu),
C(\phi) ),
\eqn\moment
$$
for all $\phi \in \Omega^{0}(X,{\bf g}_E)$, and $C$ is
the map given in \flor. The brackets denote the dual pairing. The explicit
expression of this map is given by:
$$ m(A,
\alpha)=F^{ij} \wedge \omega + {\omega \wedge \omega \over 2}(\alpha^i
{\overline \alpha}^j-{\delta^{ij} \over 2}|\alpha^k|^2).
\eqn\map
$$
The first piece of this map is just the corresponding map for Donaldson
theory, and the second piece contains the dependence on the monopole part.
The property \moment\ is easily verified from the expression for the
differential of \map:
$$
(dm)_{(A, \alpha)}(\psi, \mu)=(d_A \psi)^{ij} \wedge \omega+{\omega \wedge
\omega
\over 2}\Big(\mu^i {\overline \alpha}^j+\alpha^i {\overline
\mu}^j-{\delta^{ij} \over 2}(\mu^k {\overline \alpha}^k+\alpha^k {\overline
\mu}^k) \Big).
\eqn\luisa
$$

The solutions of the second equation in \kal\ are precisely the
zeroes of the moment map \map, as it happens in the abelian case. This
indicates that the moduli space of solutions of the $SU(2)$ monopole
equations with $\beta=0$ can be identified with the symplectic quotient
$m^{-1}(0)/{\cal G}$. Furthermore, under certain stability conditions this
symplectic quotient can be identified with what is called the complex
quotient of ${\cal M}_{\beta =0}$, {\it i.e.}, the quotient by the
complexification of the group of gauge transformations ${\cal G}^c$, which in
this case is simply the group of $Sl(2,C)$ gauge transformations. More
precisely, in order to identify the symplectic quotient with the complex
quotient one must get rid in the former of the points of the moduli space in
which the group ${\cal G}$ has a non-free action. As we have discussed in the
preceding section, as far as $M \not= 0$ the group of gauge transformations
acts freely, and therefore we don't need any additional restriction when
considering ${\cal M}_{\beta =0}$. Now recall that, because of \hol, the
connections in $A$ define holomorphic structures on the bundle $E$, and it
can be seen that two connections define isomorphic holomorphic structures if
and only if they are related by a complex gauge transformation. Then we can
identify equivalence classes of connections under the complexified gauge
group with equivalence classes of holomorphic $Sl(2,C)$ bundles.

Concerning the stability conditions for the complex quotient, although an
accurate treatment requires the analysis of the gradient flow lines
associated to the moment map, it seems that, when $M \not= 0$, there are no
topological restrictions on the holomorphic structures of the bundles. This
is already the case in the abelian theory [\mfm] and is in contrast with
Donaldson \REF\donstab{S.K. Donaldson\journal\plms&53(85)1} theory [\donstab,
\donkron], where a ${\cal G}^c$ orbit contains an irreducible ASD connection
if and only if the holomorphic bundle $E$ verifies a certain
algebro-geometric condition. Therefore, on a compact K\"ahler, spin manifold,
the moduli space of solutions to the $SU(2)$ monopole equations has three
branches: the first one corresponds to the irreducible ASD connections with
$M=0$, and can be identified with the equivalence classes of stable
holomorphic $Sl(2,C)$ bundles $E$. The second branch corresponds to pairs
consisting of an equivalence class of holomorphic $Sl(2,C)$ bundles $E$
together with a holomorphic section of $K^{1/2}\otimes E$ modulo $Sl(2,C)$
gauge transformations (the case $\alpha \not= 0$, $\beta = 0$ discussed
before). The third branch is similar to the second branch, but now
$\alpha=0$, $\beta \not= 0$, and consequently we must
consider instead holomorphic
sections of $K^{1/2}\otimes {\overline E}$. The structure of this moduli
space has obvious similarities with the abelian case and could be refined
along the same lines, but it strongly suggests that the non-abelian monopole
theory has a richer content than Donaldson theory, as one would expect from a
highly non-trivial coupling of the topological Yang-Mills multiplet to
topological matter.

\chapter{The Topological Action.}

In this section we will build the non-abelian generalization of the
topological action presented in [\toplag]. As in that case, we will use the
Mathai-Quillen formalism [\mq]. This formalism is very well suited for our
purposes since it provides a procedure to construct the action of a
topological quantum field theory starting from a moduli problem formulated in
purely geometrical terms. Indeed, we will apply it to the moduli problem
discussed in sect. 2.

The Mathai-Quillen form is essentially an adequate representative of the Thom
class of the bundle ${\cal E}$. As we discussed in sect. 2, when we integrate
over the space ${\cal M}/{\cal G}$ the pullback of this Thom class under a
section
$s$ of
${\cal E}$ we obtain the Euler characteristic of ${\cal E}$. In addition,
because of its localization properties, we can use this pullback to compute
intersection numbers in the moduli space constituted by the zeroes of $s$.
{}From the field-theory point of view, as the pullback of the Thom class
corresponds to exp$-S$, the Euler characteristic can be interpreted as the
partition function of the topological field theory, and the intersection
numbers as topological correlation functions. The Mathai-Quillen form is
constructed making use of a connection defined on
${\cal E}$. For the case in which the space ${\cal M}$ has a ${\cal
G}$-invariant metric defined on it there is a natural way to construct it as
follows [\aj]: consider on the principal bundle ${\cal M} \too {\cal M}/{\cal
G}$ the connection defined by declaring the horizontal subspaces to be the
orthogonal ones to the vertical subspaces. The latter are just the gauge
orbits given by the action of the group ${\cal G}$. This connection on the
principal bundle ${\cal M}$ induces a connection on the associated bundle
${\cal E}$ in the standard way, and this is just the connection that one
needs to construct the Mathai-Quillen form.

With the help of the connection which has been introduced we are now in the
position to write down the Mathai-Quillen form. We will use the Cartan model
for the equivariant cohomology which gives the BRST symmetry of the theory.
Hence we will deal with the Cartan model of the Mathai-Quillen form. This is
an equivariant differential form of the fibre ${\cal F}$ which can be written
as:
$$
 U={\rm e}^{-|x|^2} \int D \chi {\rm exp}\Big( {1 \over 4} \langle
\chi , \Omega \chi \rangle  + i\langle dx , \chi \rangle \Big).
\eqn\mathai
$$
In this expression, $x$ denotes a (commuting) vector coordinate for the
fibre ${\cal F}$,  $\chi$ a Grassman coordinate and the bracket a  ${\cal
G}$-invariant metric on ${\cal F}$. $\Omega$ is the universal curvature
which acts on the fibre according to the action of the group ${\cal G}$.
Now, in order to obtain a differential form on the base space ${\cal
M}/{\cal G}$ we must use the Chern-Weil homomorphism which has the effect of
substituting $\Omega$ by the actual curvature on ${\cal M}$ and thus gives a
basic differential form on ${\cal M} \times {\cal F}$. However, in the
Cartan model, due to the relation between the Cartan model and the Weil
model for equivariant cohomology, one needs to make an horizontal projection
in order to obtain a closed form on ${\cal E}$. In other words, the
differential form on ${\cal M} \times {\cal F}$ must be evaluated on the
horizontal subspace of ${\cal M}$. Once we do that, we have a form on ${\cal
E}$ which descends to a form on ${\cal M}/{\cal G}$ by simply taking the
pullback by the section ${\hat s}$. This has the effect of substituting the
coordinate $x$ by the section ${\hat s}$.

Let us describe in detail how to construct the connection on ${\cal M}$ and
how to enforce the horizontal projection. The gauge orbits are given by the
image of the map $C$ introduced in \ope. Consider now the operator
$R=C^{\dagger}C$. The connection one-form is given by [\aj],   $$
\Theta = R^{-1} C^{\dagger}.
\eqn\lima
$$
As the
Cartan representative acts on horizontal vectors, we can write the curvature
as
$$
\Omega = d \Theta =R^{-1}dC^{\dagger}.
\eqn\curv
$$
Now, to enforce the horizontal projection we should have to integrate over
the vertical degrees of freedom which amounts to an integration over the Lie
group. Alternatively, we can introduce a ``projection form" [\cmrlect] which,
besides of projecting on the horizontal direction, automatically involves the
Weil homomorphism which substitutes the universal curvature by the actual
curvature on the bundle \curv. The projection form also allows to write the
correlation functions on the quotient moduli space ${\cal M}/{\cal G}$ as
integrals over ${\cal M}$, in such a way that we can consider the original
section $s$ instead of ${\hat s}$. Taking into account all these facts, and
after some suitable manipulations, we obtain the following expression for the
Thom class of the bundle ${\cal E}$:
$$
\int  D\eta D\chi D\phi D \lambda \,
{\rm exp}\Big( -|s|^2 + {1 \over 4} \langle \chi , \phi \chi \rangle +
i\langle ds , \chi \rangle +i{\langle dC^{\dagger}, \lambda \rangle }_{g}-
i{\langle \phi,R \lambda \rangle }_{g} +i\langle C^{\dagger} \theta , \eta
\rangle_{g} \Big).
\eqn\mqaj
$$
Here, $\phi$, $\lambda$ are conmuting Lie
algebra variables and $\eta$ is a Grassmann one. The variables $(P, \theta)$
(the first one is conmuting and present in $s$, the second one is Grassmann)
are the usual superspace coordinates for the integration of differential
forms on ${\cal M}$. The bracket with the subscript $g$ is the
Cartan-Killing form of ${\rm Lie}({\cal G})$. This expression is to be
understood as a differential form on ${\cal M}$ which when integrated out
with the measure $DP D\theta$ gives the Euler characteristic of ${\cal E}$.
This ends our brief introduction of the Mathai-Quillen formalism.

We will apply the previous formalism to the  moduli problem of non-abelian
monopoles on four-manifolds introduced in sect. 2.  We will restrict
ourselves to the gauge group
$SU(N)$ but a similar construction holds in the general case. The operators
$C$ and $C^{\dagger}$ have been explicitly computed in \flor\ and \marga,
respectively. The operator $R=C^\dagger C$ is easily obtained:
$$
R(\phi)^{ij}=(d_A^{*}d_A \phi)^{ij} +
{1\over 2}({\overline M}^{\alpha k}\phi^{kj}M_{\alpha}^i
+{\overline M}^{\alpha j}\phi^{ik}M_{\alpha}^k) - {\delta^{ij}\over
d_R}{\overline M}^{\alpha k}\phi^{kl} M_\alpha^l,  \,\,\,\,\
\phi \in \Omega^{0}(X,{\bf g}).
\eqn\tren
$$
The other operator involved in \mqaj, $ds$, has been computed in \disco.

In order to write the topological quantum field  theory associated to the
moduli problem we must indicate the field content and the topological
symmetry. These are determined by the geometrical structure that we have been
developing. For the moduli space we have conmuting fields $P=(A,M) \in {\cal
M}= {\cal A} \times \Gamma (X, S^{+} \otimes E)$, with ghost number $0$ and
their superpartners, representing a basis of differential forms on ${\cal
M}$, $\theta=(\psi, \mu)$, with ghost number $1$. Now, we must introduce
fields for the fibre  which
we denote by $(\chi_{\mu\nu}, v_{\dot \alpha}) \in \Omega^{2,+} (X, {\bf
g}_E)
\oplus \Gamma (X, S^{-} \otimes E)$, with ghost number $1$. It is also
useful in the construction of the action from gauge fermions to introduce
auxiliary conmuting fields with the same geometrical content,
$(H_{\mu\nu}, h_{\dot \alpha})$. The gauge symmetry makes necessary to
introduce three fields in ${\rm Lie}({\cal G})$, as we have remarked in
writing \mqaj. The field $\phi\in \Omega^0(X,{\bf g}_E)$, with ghost number
$2$, is a conmuting one. It roughly corresponds
 to the universal curvature and enters in the equivariant cohomology of
${\cal M}$. The fields $\lambda$ and  $\eta$, also in $\Omega^0(X,{\bf g}_E)$
but anticommuting and  with ghost
number $-2$ and $-1$, respectively, come from the projection form, as
explained in [\cmrlect]. The BRST cohomology of the model is:
$$
\eqalign{ &[Q,A]=\psi,\cr
&\{Q,\psi \}=d_A\phi,\cr
&[Q, \phi]=0\cr
&\{Q,\chi_{\mu\nu} \}=H_{\mu\nu}, \cr
&[Q,H_{\mu\nu}]=i[\chi_{\mu\nu},\phi], \cr
&[Q, \lambda]=\eta, \cr}
\qquad\qquad
\eqalign{
&[Q,M_{\alpha}^i]=\mu_{\alpha}^i, \cr
&\{Q, \mu_{\alpha}^i \}=-i\phi^{ij} M_{\alpha}^j,\cr
& \{Q, v_{\dot \alpha}^i \}=h_{\dot \alpha}^i, \cr
&[Q,h_{\dot \alpha}^i]=-i \phi^{ij} v_{\dot \alpha}^j, \cr
& \{Q, \eta \}=i[\lambda,\phi].\cr}
\eqn\pera
$$
This BRST gauge algebra closes up to a gauge transformation generated by
$-\phi$ (recall that the group acts on the fibre with $g^{-1}$).

We are now in the position to write down the action of the
theory. Let us  consider first the last five terms in the exponential of the
Thom class \mqaj,
$$
\eqalign{
-&i{\langle \phi, R\lambda \rangle }_{g} =-i\int _{X} \tr(
\lambda  \wedge *d_A^{*}d_A\phi)
-{i\over 2}\int_{X}e {\overline M}^{\alpha}\{\lambda,\phi\}
M_{\alpha},   \cr
&i\langle  (\chi , v), ds \rangle = {i \over {\sqrt2}}
\int _{X} \tr (\chi \wedge * p^{+}d_A \psi)
-{1 \over {\sqrt2}}
\int _{X}e ({\overline M}_{\alpha}{\chi}^{\alpha \beta} \mu_{\beta}-
{\bar \mu}_{\alpha}{\chi}^{\alpha \beta} M_{\beta })  \cr
&\,\,\,\,\,\,\,\,\,\,\,\,\,\,\,\,\,\,\,\,\,\,\,  +{i \over 2}\int _{X} e
(\bar v^{\dot \alpha} D_{\alpha {\dot\alpha}} {\mu}^{\alpha} -{\bar
\mu}^{\alpha } D_{\alpha {\dot\alpha}}v^{\dot \alpha}) +{1 \over 2}\int _{X}
e ({\overline M}^{\alpha }{\psi}_{\alpha {\dot \alpha}}v^{\dot\alpha }-{\bar
v}^{\dot \alpha }{\psi}_{\alpha {\dot \alpha}}M^{\alpha }),\cr
&i\langle C^\dagger (\psi,\mu),
 \eta \rangle_g   = - \int_{X} \tr
(\eta \wedge *d_A^*\psi ) +{1\over2} \int _{X} e  ({\bar \mu}^{\alpha }
\eta M_{\alpha}+{\overline M}^{\alpha } \eta \mu_{\alpha}), \cr
&{1 \over 4}{\langle (\chi , v), \phi (\chi , v) \rangle} = -{i\over
4}\int_X \tr (\chi\wedge * [\phi,\chi]) - {i \over 4} \int_{X} e
{\bar v}^{\dot\alpha } \phi v_{\dot\alpha}, \cr
&i{\langle dC^{\dagger}, \lambda \rangle}_{g}=\int_X \tr
(\lambda\wedge * [\psi,*\psi]) +
\int _{X} e  {\bar \mu}^{\alpha } \lambda \mu_{\alpha}.\cr }
\eqn\fresa
$$
The section term in \mqaj\ has been computed in
\melon, and the action resulting after adding to it all the terms in \fresa\
constitutes the field theoretical representation of the Thom class of the
bundle ${\cal E}$. This action is invariant under the transformations \pera\
once the auxiliary field $H_{\alpha\beta}$ and $h_{\dot\alpha}$ are
introduced. It can be obtained in its off-shell form using the nilpotent
transformations \pera\ (up to a gauge transformation) and an appropriate
gauge invariant gauge fermion. This approach was first used in
\REF\labper{J.M.F. Labastida and M. Pernici\journal\pl&B212(88)56} [\labper]
(for a review on subsequent developments see \REF\blth{D. Birmingham, M.
Blau, M. Rakowski and G. Thompson\journal\pre&209(91)129} [\blth]) and
reformulated in the context of the Mathai-Quillen formalism in [\cmrlect]. We
will now construct the topological action using this last point of view. In a
topological field theory with gauge symmetry there exists a localization
gauge fermion which comes directly from the Cartan model representative of
the Thom class \mqaj\ with additional auxiliary fields $H_{\alpha\beta}$ and
$h_{\dot\alpha}$. In our case, the appropriate gauge fermion turns out to be:
$$
\Psi _{\rm loc}= -i\langle (\chi,
v),s(A,M) \rangle-{1 \over 4}\langle  (\chi, v),(H,h) \rangle,
\eqn\tigre
$$
while the projection gauge fermion, which implements the horizontal
projection, is,
$$
\Psi _{\rm proj}=i{\langle \lambda , C^{\dagger} (\psi ,
\mu ) \rangle}_{g}.
\eqn\oso
$$

Making use of the $Q$-transformations \pera\ one easily computes  the
localization  and the projection lagrangians:
$$
\eqalign{ \{Q,\Psi _{\rm loc}\} =
&\Big\{ Q, \int _{X} e \Big[-i\chi^{\alpha \beta ji}\Big({1 \over {\sqrt 2}}
\big(F^{+ij}_{\alpha \beta} +{i \over 2}({\overline M}_{(\alpha}^j
M_{\beta)}^i -{\delta^{ij}\over d_R}{\overline M}_{(\alpha}^k
M_{\beta)}^k)\big) - {i \over 4} H_{\alpha \beta} \Big)\cr
&\,\,\,\,\,\,\,\,\,\,\,\,\,\,\,\,\,\,\,\,\,\,\,\,\,
 -{i\over 2} ({\bar v}^{\dot \alpha}D_{\alpha \dot \alpha}
M^{\alpha}+{\overline M}^{\alpha}D_{\alpha \dot \alpha}v^{\dot \alpha})-{1
\over 8} ({\bar v}^{\dot \alpha} h_{\dot \alpha}-{\bar h}_{\dot
\alpha}v^{\dot \alpha} ) \Big] \Big\}\cr =  & \int _{X}e \Big[-{i \over
{\sqrt 2}}
 H^{\alpha \beta ji} \big(F^{+ij}_{\alpha \beta} +{i \over 2}({\overline
M}_{(\alpha}^j M_{\beta)}^i -{\delta^{ij}\over d_R}{\overline M}_{(\alpha}^k
M_{\beta)}^k)\big) \cr  &\,\,\,\,\,\,\,\,\,\,\,\,\,\,
+{i \over {\sqrt 2}}\tr\big(\chi ^{\alpha
\beta}(p^+(d_A\psi))_{\alpha\beta}\big)
+{1\over \sqrt{2}}({\bar \mu}_{\alpha}\chi^{\alpha\beta} M
_{\beta}- {\overline M}_{\alpha}\chi^{\alpha\beta}\mu _{\beta})
 \cr
&\,\,\,\,\,\,\,\,\,\,\,\,\,
-{1 \over 4} \tr(H^{\alpha \beta} H_{\alpha \beta})
+{i\over 4} \tr(\chi^{\alpha\beta}[\chi_{\alpha\beta},\phi])
-{i\over 2} ({\bar h}^{\dot \alpha}D_{\alpha \dot \alpha}
M^{\alpha}+{\overline M}^{\alpha}D_{\alpha \dot \alpha}h^{\dot \alpha})
\cr
&\,\,\,\,\,\,\,\,\,\,\,\,\,\,
+{i \over 2}({\bar v}^{\dot \alpha}D_{\alpha \dot \alpha} \mu^{\alpha}-{\bar
\mu}^{\alpha}D_{\alpha \dot \alpha}v^{\dot \alpha}) + {1 \over 2}
 ({\overline M}^{\alpha} \psi_{\alpha \dot
\alpha}v^{\dot \alpha}- {\bar v}^{\dot \alpha} \psi_{\alpha \dot
\alpha}M^{\alpha})\cr
&\,\,\,\,\,\,\,\,\,\,\,\,\,\,
-{1 \over 4} ({\bar h}^{\dot\alpha} h_{\dot\alpha}
+i{\bar v}^{\dot\alpha} \phi v_{\dot\alpha}) \Big],\cr}
\eqn\manza
$$
$$
\eqalign{ \{Q,\Psi _{\rm proj}\}= &\{ Q, -\int _{X} \big[i
 \tr(\lambda \wedge *d_A^{*} \psi) +{1 \over 2} e  ({\bar
\mu}^{\alpha}\lambda M_{\alpha}- {\overline M}^{\alpha} \lambda
\mu_{\alpha}) \big] \}\cr  =
&\int _{X} \big[\tr\big(- i\eta \wedge *d_A^{*} \psi
-i \lambda \wedge *d_A^{*}d_A \phi - \lambda\wedge *[*\psi,\psi] \big)
\cr  & \,\,\,\,\,\,\,
+ {1 \over 2} e  ({\bar \mu}^{\alpha}\eta M_{\alpha} + {\overline M
}^{\alpha} \eta \mu_{\alpha}) + e  ({\bar
\mu}^{\alpha}\lambda\mu_{\alpha} -{i\over 2} {\overline M}^{\alpha}
\{\phi,\lambda\} M_{\alpha} \big)\big]. \cr}  \eqn\limon
$$

The sum of \manza\ and \limon\ is just the same as the
sum of the terms in \fresa\ plus $-|s(A,M)|^2$ as given in \melon\
once the auxiliary fields
$H_{\alpha\beta}$ and $h_{\dot\alpha}$ have been integrated out.
This is indeed the exponent appearing in the Thom class \mqaj\ which must
be identified as minus the action, $-S$, of the topological quantum field
theory. After carrying out the integration of the auxiliary fields the
resulting action turns out to be:
$$
\eqalign{S=&
\int_{X} e \big[ g^{\mu\nu}D_\mu\overline M^\alpha D_\nu
M_\alpha + {1\over 4} R \overline M^\alpha M_\alpha +{1\over 2}
\tr (F^{+\alpha\beta} F_{\alpha\beta}^+)
-{1\over 8}(\overline
M^{(\alpha} T^a M^{\beta)}) (\overline M_{(\alpha} T^a M_{\beta)})]
\cr
+&\int_{X} \tr\big(\eta \wedge *d_A^{*} \psi
-{i \over {\sqrt 2}}\chi ^{\alpha
\beta}(p^+(d_A\psi))_{\alpha\beta}
-{i\over 4}\chi^{\alpha\beta}[\chi_{\alpha\beta},\phi]
+ i \lambda \wedge *d_A^{*}d_A \phi + \lambda\wedge *[*\psi,\psi] \big)
\cr
+&\int_{X}e\Big(-i{\overline M}^{\alpha}\{\phi, \lambda\} M_{\alpha}  +{1
\over {\sqrt2}} ({\overline M}_{\alpha} {\chi}^{\alpha \beta} \mu _{\beta
}-{\bar \mu}_{\alpha}{\chi}^{\alpha \beta} M_{\beta}) -{i\over 2} (v^{\dot
\alpha} D_{\alpha {\dot\alpha}} {\mu}^{\alpha}-{\bar \mu}^{\alpha} D_{\alpha
{\dot\alpha}}v^{\dot \alpha})  \cr
&\,\,\,\,\,\,\,\,\,\,\,\,\,\,\,\,\,\,-{1 \over 2}({\overline
M}^{\alpha}{\psi}_{\alpha {\dot \alpha}}v^{\dot \alpha}-{\bar v}^{\dot
\alpha}{\psi}_{\alpha {\dot \alpha}}M^{\alpha}) -{1\over2}  ({\bar
\mu}^{\alpha}  \eta M_{\alpha}+{\overline M}^{\alpha}
 \eta\mu_{\alpha}) +{i \over 4}{\bar v}^{\dot\alpha} \phi v_{\dot\alpha}
-{\bar \mu}^{\alpha} \lambda \mu _{\alpha}\Big),\cr }
\eqn\action
$$
where we have used \current\ to write the term quartic in the
fields $M_\alpha$. Although this action has been computed considering an
arbitrary representation of the gauge group $SU(N)$, its form is also valid
for any other gauge  group. This action is
invariant under the modified BRST transformations which are obtained from
\pera\ after taking into account the modifications which appear once
the auxiliary fields have been integrated out. It contains the standard
gauge fields of a twisted $N=2$ vector multiplet, or Donaldson-Witten
fields, coupled to the matter fields of the twisted $N=2$ hypermultiplet.

One important question is the analysis of the observables of the theory.
Certainly one has the observables corresponding to ordinary Donaldson theory.
These have been written down explicitly for the abelian case in [\toplag].
For the non-abelian case these are the ones in [\tqft] and their
generalization for an arbitrary gauge group.  The issue now is to study if
there are some observables involving matter fields, \ie, BRST invariant
operators which are not $Q$-exact. We have not found any.

\chapter{Conclusions.}

In this paper we have presented the non-abelian generalization of
Witten's monopole equations. These equations lead to the study of a new
moduli problem which is a generalization of Donaldson theory.
We have performed a first analysis of the space of solutions and
it has been argued that they constitute an enlarged moduli space which might
be richer than the ordinary one. In addition, the paper contains the
formulation of  the topological quantum field theory associated to the moduli
problem corresponding to the non-abelian monopole equations. This has been
carried out using the Mathai-Quillen formalism and the resulting theory
contains twisted gauge and matter $N=2$ supersymmetric multiplets.

This work opens a variety of investigations. Certainly, the moduli space of
solutions should be further analyzed. Our study can be consider as a
preliminary one which seems to suggests that there exist an interesting
structure, but a refined analysis should be performed.
Another important study which should be carried out is the analysis of the
moduli problem presented in this paper from the point of
view of the underlying untwisted $N=2$ theory. In particular, it would be
interesting to know if the topological quantum field theory
can be analyzed  along the lines of [\wjmp]  for the K\" ahler case,
or using the
techniques of [\ws] in the general case. One could ask then if there exist a
theory with a moduli problem equivalent to the one presented in this work
similarly as it happens between ordinary Donaldson theory and the abelian
monopole equations.

Another important aspect of the topological quantum field theory presented
in this work is the study of its possible relation to string theory
following the type of analysis done in \REF\hast{J.A. Harvey and A.
Strominger\journal\cmp&151(93)221} [\hast] for the case of ordinary Donaldson
theory.

\vskip1cm

\ack We would like to thank A. V. Ramallo for very helpful discussions.
M.M. would like to
thank G. Moore and S. Rangoolam for many conversations about the
Mathai-Quillen formalism.
This work was supported in part by DGICYT under grant PB93-0344 and
by CICYT under grant
AEN94-0928.

\vskip1cm

{\bf Note added:} After this work was completed we became aware of
\REF\park{S. Hyun, J. Park and J.S. Park, ``Topological
QCD", hep-th/9503201} [\park] where a topological quantum field theory
similar to ours is considered.

\endpage

\appendix

In this appendix we summarize the conventions used in this paper. Basically
we will describe the elements of the positive and
negative chirality spin bundles $S^+$ and $S^-$ on a four-dimensional spin
manifold $X$ endowed with a vierbein $e^{m\mu}$ and a spin connection
$\omega_{\mu}^{m n}$.  Let us begin recalling that a Dirac spinor in
Euclidean four-dimensional space corresponds to $S^+\oplus S^-$ and it is
associated to a representation of dimension four of the group of rotations
$SO(4)$. This representation is reducible in terms of the simplest
irreducible representation of $SO(4)$: the one associated to two-component
Weyl spinors. These describe locally the elements of $\Gamma(X,S^+)$ and
$\Gamma(X,S^-)$ on the spin manifold $X$. The four-dimensional representation
of $SO(4)$ assocaited to $S^+\oplus S^-$ is built out of gamma matrices
satisfying: $$ \{\gamma_m,\gamma_n\}=2 \delta_{mn}.
\eqn\launo
$$
These can be chosen to be hermitian:
$$
\gamma_0 = \left(\matrix{0&{\bf 1}\cr
                         {\bf 1}&0\cr}\right),\,\,\,\,\,\,\,\,\,\,\,
\gamma_a =\left(\matrix{0&{i\tau_a}\cr
                         {-i\tau_a}&0\cr}\right),
\,\,\,\,\,\,\, a=1,2,3,
\eqn\lasgammas
$$
where ${\bf 1}$ is the $2\times 2$ unit matrix and
$\tau_a$, $a=1,2,3$ are the Pauli matrices:
$$
\tau_1 = \left(\matrix{0&{1}\cr
                         {1}&0\cr}\right),\,\,\,\,\,\,\,\,\,\,\,
\tau_2 = \left(\matrix{0&{-i}\cr
                         {i}&0\cr}\right),\,\,\,\,\,\,\,\,\,\,\,
\tau_3 = \left(\matrix{1&{0}\cr
                         {0}&-1\cr}\right).
\eqn\lados
$$
Throughout this appendix latin indices at the beginning of the alphabet
$a,b,\cdots$ will run from 1 to 3 (unless otherwise indicated),
while the ones at the middle $m,n,\cdots$ will run from 0 to 3.
The Pauli matrices satisfy: $$
\tau_a\tau_a = i\epsilon_{abc}\tau_c+\delta_{ab}{\bf 1},
\eqn\latres
$$
where $\epsilon_{abc}$ is the totally antisymmetric tensor with
$\epsilon_{123}=1$. The projection from a Dirac four-component spinor to a
Weyl two-component one is carried out with the help of the matrix $\gamma_5$
which verifies $\{\gamma_5,\gamma_m\}=0$ and is chosen to be:
$$
\gamma_5 = \left(\matrix{{\bf 1}&{0}\cr
                         {0}&-{\bf 1}\cr}\right).
\eqn\gammafive
$$
The projection into the positive and negative chirality spin bundles
$S^+$ and $S^-$ is performed using ${1\over 2}(1+\gamma_5)$ and ${1\over
2}(1-\gamma_5)$ respectively.

The four-dimensional representation of $SO(4)$ associated to
$S^+\oplus S^-$ has the following matrices:
$$
\gamma_{mn} = {i\over 4} [\gamma_m,\gamma_n],
\eqn\lacuatro
$$
which have been chosen to be hermitian.
They satisfy the $SO(4)$ algebra:
$$
[\gamma_{mn},\gamma_{pq}]=
i\big(\delta_{np}\gamma_{mq}
-\delta_{nq}\gamma_{mp}
+\delta_{mp}\gamma_{qn}
-\delta_{mq}\gamma_{pn}\big).
\eqn\lacinco
$$

Using the representation \lasgammas\ and \lacuatro\ one immediately obtains
the two two-dimensional representations of $SO(4)$ which correspond to
$S^+$ and $S^-$. Indeed, the matrices $\gamma_{mn}$ have the form:
$$
\gamma_{mn} = \left(\matrix{\sigma_{mn}&{0}\cr
                         {0}&\tilde\sigma_{mn}\cr}\right),
\eqn\laseis
$$
where the matrices $\sigma_{mn}$ and $\tilde\sigma_{mn}$
are antisymmetric in $m$ and $n$ and have the form,
$$
\eqalign{
&\sigma_{0a}={1\over 2}\tau_a,\cr
&\tilde\sigma_{0a}=-{1\over 2}\tau_a, \cr}
\qquad\qquad
\eqalign{
&\sigma_{ab}=-{1\over 2}\epsilon_{abc}\tau_c, \cr
&\tilde\sigma_{ab}=-{1\over 2}\epsilon_{abc}\tau_c. \cr}
\eqn\lasiete
$$
Certainly, from \lacinco\ and \laseis\ follows that the matrices
$\sigma_{mn}$ and $\tilde\sigma_{mn}$ satisfy the $SO(4)$ algebra.
Furthermore, the matrices of the two sets are hermitian.

Under an infinitesimal $SO(4)$ rotation a Weyl spinor $M^\alpha$,
$\alpha=1,2$, associated to $S^+$, transforms as:
$$
\delta M^\alpha = {1\over 2} \epsilon_{mn}
(\sigma_{mn})^\alpha{}_\beta M^\beta,
\eqn\laocho
$$
where $\epsilon_{mn}=-\epsilon_{nm}$ are the infinitesimal parameters
of the transformation. On the other hand, a Weyl spinor $N_{\dot\alpha}$,
$\dot\alpha=1,2$, associated to $S^-$, transforms as,
$$
\delta N_{\dot\alpha} = {1\over 2} \epsilon_{mn}
(\tilde\sigma_{mn})_{\dot\alpha}{}^{\dot\beta} M_{\dot\beta}.
\eqn\lanueve
$$
The Pauli matrix $(\tau_2)_{\alpha\beta}$ is an invariant tensor as one can
easily verify:
$$
(\sigma_{mn})_\alpha{}^\gamma (\tau_2)_{\gamma\beta}+
(\sigma_{mn})_\beta{}^\gamma (\tau_2)_{\alpha\gamma}=
(\sigma_{mn}\tau_2 + \tau_2\sigma_{mn}^\top)_{\alpha\beta}=0,
\eqn\invariancia
$$
after using the fact that since
$$
\tau_2\tau_a^\top\tau_2=-\tau_a,\,\,\,\,\,\,\, a=1,2,3,
\eqn\latreinta
$$
one has,
$$
\tau_2 \sigma_{mn}^\top \tau_2 = - \sigma_{mn},
\,\,\,\,\,\,\,\,\,\,\,\,\,\,
\tau_2 \tilde\sigma_{mn}^\top \tau_2 = - \tilde\sigma_{mn}.
\eqn\nasinva
$$

The invariant matrix $\tau_2$  can be
used to raise and lower spinor indices. Following the conventions in
\REF\roc{S.J. Gates, M.T. Grisaru, M. Ro\v cek
and W. Siegel, ``Superspace", Benjamin, 1983} [\roc] we define:
$$
C_{\alpha\beta} =  (\tau_2)_{\alpha\beta},
\,\,\,\,\,\,\,\,\,\,\,\,\,\,\,
C_{\dot\alpha\dot\beta} =  (\tau_2)_{\dot\alpha\dot\beta},
\eqn\lasces
$$
and their inverse tensors:
$$
C^{\alpha\beta} = - (\tau_2)_{\alpha\beta},
\,\,\,\,\,\,\,\,\,\,\,\,\,\,\,
C^{\dot\alpha\dot\beta} = - (\tau_2)_{\dot\alpha\dot\beta},
\eqn\lascesmas
$$
so that,
$$
C^{\alpha\beta}C_{\gamma\beta} = \delta_\gamma^\alpha,
\,\,\,\,\,\,\,\,\,\,\,\,\,\,\,
C^{\dot\alpha\dot\beta}C_{\dot\gamma\dot\beta} =
\delta_{\dot\gamma}^{\dot\alpha},
\eqn\lascesmasmas
$$
and therefore can be utilized to raise and lower spinor indices:
$$
\eqalign{
M_\alpha &= M^\beta C^{\beta\alpha} , \cr
M^\alpha &= C^{\alpha\beta} M_\beta, \cr }
\qquad\qquad
\eqalign{
N^{\dot\alpha} &= C^{\dot\alpha\dot\beta} N_{\dot\beta}, \cr
N_{\dot\alpha} &= N^{\dot\beta }C^{\dot\beta\dot\alpha}. \cr}
\eqn\laonce
$$

It is useful to study how $M^\alpha$ and $N_{\dot\alpha}$ transform
under infinitesimal $SO(4)$ rotations. One finds in this way another two
realizations of the two-dimensional representation. Using \laocho\ and
\laonce\ one finds:
$$
\eqalign{
\delta M_\alpha &= {1\over 2} \epsilon_{mn}
(\sigma_{mn}')_\alpha{}^\beta M_\beta,\cr
\delta N^{\dot\alpha} &= {1\over 2} \epsilon_{mn}
(\hat\sigma_{mn})^{\dot\alpha}{}_{\dot\beta} N^{\dot\beta},\cr}
\eqn\latrece
$$
where,
$$
\sigma_{mn}'=\tau_2\sigma_{mn}\tau_2=-\sigma_{mn}^\top,
\,\,\,\,\,\,\,\,\,\,\,\,
\hat\sigma_{mn}=\tau_2\tilde\sigma_{mn}\tau_2=-\tilde\sigma_{mn}^\top.
\eqn\lacatorce
$$

In order to write down the Dirac equation for a Weyl spinor we need to
introduce new matrices. Let us define the set of four matrices $\sigma_m$,
$m=0,1,2,3$, as:
$$
\sigma_0 = {\bf 1}, \,\,\,\,\,\, \sigma_a=i\tau_2\tau_a\tau_2
=-i\tau_a^\top, \,\,\,\,\,
a=1,2,3. \eqn\ladoce
$$
This is a convenient choice because on the  one hand the determinant
of $P_m\sigma_n$ is,
$$
\det [P_m\sigma_n] = P_0^2+P_1^2+P_2^2+P_3^2,
\eqn\latrece
$$
and, on the other hand, it transforms as a vector under $SO(4)$ rotations.
Indeed, one has,
$$
{i\over 2}\epsilon_{mn}(\sigma_{mn})_\alpha{}^\beta
(\sigma_p)_{\beta\dot\alpha}
+{i\over 2}\epsilon_{mn}(\hat\sigma_{mn})_{\dot\alpha}{}^{\dot\beta}
(\sigma_p)_{\alpha\dot\beta}={1\over 2}\epsilon_{mn}
\delta_{p[m}(\sigma_{n]})_{\alpha\dot\alpha},
\eqn\laveinte
$$
where $\sigma_{mn}$ and $\hat\sigma_{mn}$ are given in \lasiete\
and \lacatorce\ respectively.

Let us consider the covariant derivative ${\cal D}_\mu$ on the manifold
$X$. Acting on an element of $\Gamma(X,S^+)$ it has the form:
$$
{\cal D}_\mu M_\alpha = \partial_\mu M_\alpha - {i\over 2}
\omega_{\mu}^{m n} (\sigma_{mn})_\alpha{}^\beta M_\beta,
\eqn\cova
$$
where $\omega_{\mu}^{m n}$ is the spin connection. Defining
${\cal D}_{\alpha\dot\alpha}$ as,
$$
{\cal D}_{\alpha\dot\alpha} = (\sigma_n)_{\alpha\dot\alpha} e^{n\mu}
{\cal D}_\mu,
\eqn\mascova
$$
where $e^{n\mu}$ is the vierbein on $X$, the Dirac equation for
$M\in \Gamma(X,S^+)$ and $N\in \Gamma(X,S^-)$
can be simply written as,
$$
{\cal D}_{\alpha\dot\alpha} M^\alpha=0,
\,\,\,\,\,\,\,\,\,\,\,\,\,
{\cal D}_{\alpha\dot\alpha} N^{\dot\alpha}=0.
\eqn\dirac
$$
Explicit computations show that these equations are equivalent to
$$
{1\over 2}e^{m\mu}\gamma_\mu(1\pm\gamma_5)\Psi=0,
\eqn\diracfull
$$
where $\Psi\in\Gamma(X,S^+\oplus S^-)$ is the Dirac spinor,
$$
\Psi = \left(\matrix{M^\alpha\cr
          N_{\dot\alpha}\cr}\right).
\eqn\diracspinor
$$

Let us now introduce a $G$ gauge connection $A$ and let us consider
that the Weyl spinors $M_\alpha^i$ realize locally an element of
$\Gamma(S^+\otimes E)$, \ie, they transform under an $G$ gauge
transformation in a representation $R$:
$$
\delta M_\alpha^i = i\phi^{ij} M_\alpha^j = i\phi^a (T^a)^{ij}
M_\alpha^j,
\eqn\gaugetransf
$$
where $T^a$, $a=1,\cdots,N^2-1$  are the
generators of $G$ in the representation $R$, which are traceless
and chosen to be hermitian. In \gaugetransf\ $\phi^a$, $a=1,\cdots,N^2-1$,
denote the infinitesimal parameters of the gauge transformation. We use
the same type of indices as in \latres\ to label the group generators but
there is not risk to be mistaken because their meaning will be always clear
from the context. Using the gauge connection $A$ and taking \cova\ we define
now the full covariant derivative,
$$
{D}_\mu M_\alpha^i = \partial_\mu M_\alpha^i - {i\over 2}
\omega_{\mu}^{m n} (\sigma_{mn})_\alpha{}^\beta M_\beta^i
+i A_\mu^{ij}M_\alpha^j,
\eqn\covados
$$
and its analogue in \mascova:
$$
{D}_{\alpha\dot\alpha} = (\sigma_n)_{\alpha\dot\alpha} e^{n\mu} {D}_\mu.
\eqn\mascovados
$$
In the context under consideration the Dirac equations \dirac\ become:
 $$
{D}_{\alpha\dot\alpha} M^{\alpha i}=0,
\,\,\,\,\,\,\,\,\,\,\,\,\,
{D}_{\alpha\dot\alpha} N^{\dot\alpha i}=0.
\eqn\diracdos
$$

Given an element of $\Gamma(X,S^+\otimes E)$, $M_\alpha^i = (a^i,b^i)$ we
define $\overline M^{\alpha i} = (a^{i*},b^{i*})$ to be the corresponding
one of $\Gamma(S^+\otimes \overline E)$ where $\overline E$ denotes
the bundle associated
to the representation conjugate to $R$. In this way, given $M,N \in \Gamma(X,
S^+\otimes E)$, the gauge-invariant quantity entering the metric \pina,
$$
\overline M^{\alpha i} N_{\alpha}^i +
\overline N^{\alpha i} M_{\alpha}^i,
\eqn\positiva
$$
is positive definite.

Acting on an element of $\Gamma(X,S^+\otimes E)$ the covariant derivatives
satisfy:
$$
[D_\mu,D_\nu]M_\alpha^i = i F_{\mu\nu}^{ij} M_\alpha^j +
i R_{\mu\nu}{}^{mn}(\sigma_{mn})_\alpha{}^\beta M_\beta^i,
\eqn\conmutador
$$
where $F_{\mu\nu}^{ij}$ are the components of the two-form field strength:
$$
F = d A + A\wedge A,
\eqn\fuerzacampo
$$
and $R_{\mu\nu}{}^{mn}$ the components of the curvature two-form,
$$
R^{mn} = d\omega^{mn}+\omega^{mp}\wedge \omega^{pn},
\eqn\curvatura
$$
being $\omega^{mn}$ the spin connection one-form. The scalar curvature is
defined as:
$$
R=e^{\mu}_m e^{\nu}_n R_{\mu\nu}{}^{mn}.
\eqn\curvescalar
$$

The two-form $F$ can be decomposed into its self-dual and anti-self-dual
parts:
$$
F^\pm = {1\over 2} (F \pm *F),
\eqn\masmenos
$$
or, in components, after defining,
$$
F_{\alpha\dot\alpha,\beta\dot\beta}^{ij}=
(\sigma_m)_{\alpha\dot\alpha} (\sigma_n)_{\beta\dot\beta}
e^{m\mu} e^{n\nu} F_{\mu\nu}^{ij}=
C_{\alpha\beta} F^{-ij}_{\dot\alpha\dot\beta}+
C_{\dot\alpha\dot\beta} F^{+ij}_{\alpha\beta},
\eqn\decomp
$$
as,
$$
\eqalign{
F^{+ij}_{\alpha \beta}& = (p^+(F))_{\alpha\beta}^{ij}=e^{m\mu} e^{n\nu}
C^{\dot\alpha\dot\beta}(\sigma_m)_{\alpha\dot\alpha}
(\sigma_n)_{\beta\dot\beta}
{1\over 2}(F_{\mu\nu}^{ij}+{1\over
2e}\epsilon_{\mu\nu}{}^{\rho\sigma}F_{\rho\sigma}^{ij}),\cr
F^{-ij}_{\dot\alpha \dot\beta}& =
(p^-(F))_{\dot\alpha\dot\beta}^{ij}=e^{m\mu} e^{n\nu}
C^{\alpha\beta}(\sigma_m)_{\alpha\dot\alpha}
(\sigma_n)_{\beta\dot\beta} {1\over 2}(F_{\mu\nu}^{ij}-{1\over
2e}\epsilon_{\mu\nu}{}^{\rho\sigma}F_{\rho\sigma}^{ij}),\cr}
 \eqn\proy
$$
where the projector $p^\pm$ has been introduced.
{}From their definition, the components $F^{+ij}_{\alpha \beta}$ and
$F^{-ij}_{\dot\alpha \dot\beta}$ of the
self-dual and anti-self-dual parts of $F$ are symmetric in their
tangent-space  indices. From \proy\ one can verify that the quantity: $$
F^{+\alpha\beta ij}F_{\alpha\beta}^{+ji}=
\tr (F^{+\alpha\beta}F_{\alpha\beta}^+),
\eqn\rubio
$$
is positive definite.

Equation \conmutador\ can be rewritten in terms of the covariant derivatives
\mascovados\ in the following way:
$$
[D_{\alpha\dot\alpha},D_{\beta\dot\beta}]M_\gamma^i = i
F_{\alpha\dot\alpha,\beta\dot\beta}^{ij} M_\gamma^j + i
R_{\alpha\dot\alpha,\beta\dot\beta}{}^{mn}
(\sigma_{mn})_\gamma{}^\delta M_\delta^i,
\eqn\conmutadordos
$$
where $F_{\alpha\dot\alpha,\beta\dot\beta}^{ij}$
 is given in \decomp\ and,
$$
R_{\alpha\dot\alpha,\beta\dot\beta}{}^{mn}=
(\sigma_p)_{\alpha\dot\alpha} (\sigma_q)_{\beta\dot\beta}
e^{p\mu} e^{q\nu} R_{\mu\nu}{}^{mn}.
\eqn\laotraerre
$$
Using \conmutadordos\ and the fact that for arbitrary spinors
$M_\alpha$ and $N_\alpha$ one has $M_{[\alpha}N_{\beta]}=
C_{\alpha\beta} M_\gamma N^\gamma$, one finds,
$$
D_{\alpha\dot\alpha}
D_\beta{}^{\dot\alpha} M^{\beta i} = (g^{\mu\nu}D_\mu D_\nu - {1\over 4} R)
M_\alpha^i + i F^{+ij}_{\alpha\beta} M^{\beta j},
\eqn\wzk
$$
where $R$ is the scalar curvature \curvescalar.

\endpage

\refout
\end